\documentclass[12pt]{article}

\newcommand{\MSsch}{{\overline{\rm MS}}}
\newcommand{\al}{\alpha}
\newcommand{\be}{\begin{equation}}
\newcommand{\ee}{\end{equation}}
\newcommand{\ice}[1]{\relax}
\begin{document}
\begin{flushright}
INR 0903-95\\
hep-ph/9510207\\
\end{flushright}
\vspace*{0.4cm}

\vskip 1cm

\begin{center}
{\Large \bf Running coupling at small momenta, \\
renormalization schemes and renormalons}

\vskip 2cm

{\large \bf N.V. Krasnikov} and {\large \bf A.A. Pivovarov}

\vskip 1cm
{\it Institute for Nuclear Research of the Russian Academy of Sciences,

Moscow 117312, Russia}

\end{center}

\vskip 2cm
\begin{abstract}
We suggest a method of summing the 
perturbation theory (PT) asymptotic series related 
to infrared (IR) renormalons in QCD by 
using special renormalization schemes in which the running 
coupling can be integrated over the region of small momenta. 
For the method to work, one should consider higher order PT corrections
to the standard bubble-chain diagrams. High-order corrections 
allow one to choose a scheme in which the coupling
evolution can smoothly be extrapolated to small momenta.
In such schemes the sum of an (extended) 
IR-renormalon asymptotic series is defined
as an integral of the running coupling over the IR region. 
We give explicit examples of renormalization schemes in QCD which  
can be used for summing IR-renormalon asymptotic series according to
our definition.
\end{abstract}

\thispagestyle{empty} 

\newpage
High-order behavior of perturbation theory series in quantum field
theory determines the analytic structure of amplitudes  as functions
of the coupling constant. This analytic structure  encodes important
information on  the ultimate precision obtainable in the finite-order
PT analysis for physical observables.  A general approach to analyzing
high-order PT expansions is based on the saddle-point
approximation for the path integral evaluation~\cite{lip,lip1,JZ}.
The leading asymptotic behavior of the PT
coefficients has been investigated in some models of quantum field theory
while there are no exact results in QCD.  The high-order behavior  can
also be studied by selecting large contributions of some particular
subclass of diagrams \cite{thooft,lautrup}.  In QED a set of diagrams
generated by inserting an infinite number  of one-loop fermionic
corrections into the photon propagator (``bubble-chain'' diagrams) is
usually considered as a relevant one in this context. In other gauge
models (for instance, in QCD)  the selection of relevant diagrams is
more involved due to the necessity to retain gauge invariance.
The divergent series \cite{hardy}
emerging from such a type of approximation for  the high-order
behavior  -- ``bubble chains'' or renormalons -- have intensively been
studied \cite{Parisi,Mueller,bal}.   The ultraviolet (UV) renormalons
refer to asymptotic series emerging from the integration over large
energies while the infrared (IR) renormalons correspond to integration
over small energies.  Depending on the model, renormalons lead to
alternating (Borel summable) asymptotic series or non-alternating
(Borel non-summable)  asymptotic series.   It is generally assumed
that the Borel summation procedure  gives the proper  value for the
asymptotic  series which means that the theory can cope with  the
renormalon singularities that give alternating asymptotic series. The
appearance of Borel non-summable series is usually considered as a
signal that a theory (or an approximation in which it is studied) may
be inconsistent.  In QCD the IR renormalons give an example of  Borel
non-summable series if the leading-order approximation for the QCD
$\beta$-function with a positive coefficient $\beta_0$ (not too large
number of flavors) is used.

The recent interest in IR renormalons in QCD has been connected with
nonperturbative (nonPT) power corrections \cite{zakh}. This interest 
is caused
by a limiting accuracy of finite-order PT expressions that cannot
presently compete with a high precision of experimental data
(e.g. \cite{pivKraj0}); the accuracy of 
theoretical formulae has to be improved. A
regular way of going beyond a finite-order PT approximation
consists in   the
inclusion of nonPT terms as power corrections to the ordinary PT
expressions.  For the two-point correlators of interpolating hadronic
currents the nonPT terms can be represented
phenomenologically as vacuum expectation values of some local
operators \cite{politzer,SVZcharm} within the operator product
expansion (OPE) which is  a general method of quantum field theory
\cite{Wilson,david,tmf}. Such an approach has been successful in
investigating hadron properties in QCD \cite{SVZ}. However, there is
no well-established technique  to calculate nonPT contributions in
cases when OPE is not applicable. It has been suggested to consider
the IR renormalons in QCD (or, generally, Borel non-summable
asymptotic series of perturbation
theory) as a source for power corrections in such cases.

In QED, the bubble-chain diagrams form a gauge invariant class 
and give dominant contributions in the framework of
$1/N_F$ expansion \cite{Coquer}. In QCD the relevant diagrams are 
the gluon bubbles which do not form a gauge invariant set.
Therefore, to find renormalon
contributions in QCD one applies
a method of "naive
nonabelianization" to the same set of diagrams as in QED \cite{be0,kor}.
This is equivalent to
the substitution of the leading order running  coupling constant into
integrals over the phase space of the original diagram.   Thus, the
renormalon approximations are  attractive for discussing power
corrections from a practical point of view because  the relevant
diagrams are closely connected with the running of the coupling
constant within the renormalization group analysis.   The renormalon
contributions can be obtained by either a direct summation of the
selected set of diagrams or performing the integration of the
running coupling constant over the phase space of the leading order
diagram  which serves as a source for a bubble-chain generation.
These ways of obtaining the renormalon contributions  (the direct
summation of a given subclass of diagrams vs the integration of the
running coupling constant) are similar to the methods by which the
ultraviolet asymptotics of Green's functions in 
quantum electrodynamics was originally found. First the direct summation of 
the leading ultraviolet asymptotics in all orders was
performed~\cite{khal} and later the same results were obtained
by computing the one-loop $\beta$-function and solving the renormalization 
group  (RG) equation
for the invariant charge~\cite{RGeqGM,RGeqBS}. Note
that to use the direct summation in high orders of PT is technically
difficult while to solve the RG equations with a high-order
$\beta$-function is feasible.  This observation is also valid for
renormalons: to analyze corrections to the leading order renormalon
expressions within the direct summation approach is technically
difficult while the integration of the running coupling constant can
practically be done at any order of the approximation for the
evolution if the corresponding  $\beta$-function is known. The
technique for calculation of RG functions is well developed
\cite{aavlad} and the high-order PT expressions for $\beta$-functions are
available for a number of models.

In the asymptotically free QCD, the running coupling is
regular at large momenta  and integrals over large momenta are well
defined which solves the problem of summing the UV-renormalon
contributions.  In the direct summation technique these  contributions
correspond to alternating asymptotic series which are Borel summable
and can be obtained by a formal  term-by-term integration  of the
leading order PT expansion for  the running coupling.  For
determining the IR-renormalon contributions in QCD  one should
integrate the running coupling over small energies  where it is
singular in the leading order of PT  (Landau pole). Therefore, the
integral (and the contribution  of IR renormalons) is not defined in
the leading order of the coupling evolution.  By expanding
the running coupling in PT series  under the integration sign
and term-by-term  integration of the resulting series one arrives at
a divergent series which is not Borel summable.  This corresponds to
the direct summation method.  Thus, the problem of summing the
contributions stemming from  the infrared region can be reduced (at
least at the formal computational level) to defining an integral of
the running coupling over the small momenta. 
The interpretation of the integral
is necessary if there is a
singularity in the IR region as in the  leading order of PT in QCD
where there is a Landau pole.
A particular definition of the evolution  in the IR
region provides a recipe for resummation of the Borel non-summable
asymptotic series emerging from the integration over the small
momenta. Because the evolution  in the IR region is not perturbative
the extrapolation of the running to small momenta is not
unique and implies some nonPT assumptions.  For instance, the Landau
pole appears for the evolution generated  by the leading order
PT $\beta$-function.  This singularity is unstable against inclusion
of higher order PT corrections which are large  in the IR region
because it is a domain of strong coupling.  Higher order corrections
can strongly change the behavior of the running coupling 
at small energies.  For instance, the
pole can disappear (infrared fixed point) or the singularity can
become an integrable one and the integration will lead to a physically
acceptable result. In such cases the
integration is possible that defines the way of resummation of the
asymptotic series generated by the IR renormalons which means that
high-order PT corrections stabilize the theory.  An acceptance of a
particular scenario for the IR evolution, 
however, cannot be justified in PT framework only
and means that some nonPT assumptions have been made.

In the present paper we define the renormalon contributions using
the method of the integration of the running coupling.
The relevant
representation can be written in a compact form and generalized to
higher orders of PT. In the leading order it is equivalent to the
"naive nonabelianization" trick.   The IR contribution in QCD is not
defined in the leading order because of the singularity of the
formally extrapolated  running coupling to small
momenta. However, the method of integration allows one to generalize
the renormalon expressions to higher orders of  the coupling
evolution. In higher orders there is a freedom of choosing
renormalization schemes for the coupling and one can always
find a scheme in which the evolution trajectory is integrable in the
IR region \cite{krpiv}. With an integrable coupling one obtains a
particular way of resummation of renormalon-type contributions.  
In QED the generalization to higher orders
means  that one inserts 
the full one-particle irreducible polarization function
into the photon propagator.
It is close in spirit to organization of the expansion in skeleton
diagrams. In QCD, the renormalon chains beyond the leading order can be
generated from the running coupling
using Dyson-Schwinger equation \cite{kraspiv}.

Practical applications of the renormalon-type estimates are most
frequent in the standard QCD  \cite{beneke}.  An important consequence
of the bubble-chain based analysis for 
heavy quarks is an IR sensitivity of the pole mass \cite{HQm}.
Because the IR region is a nonperturbative one and 
the results based
on the consideration  of bubble chains are not rigorous
(e.g. \cite{silv,suslov})
there is no unique quantitative evaluation of such a sensitivity.
In high precision analysis of heavy quark
properties it is generally assumed that the uncertainty of the heavy
quark mass due to the IR sensitivity is numerically small
\cite{hoang,pp}.

In the following we explore some possibilities to define an integral
of the running coupling over the IR region  by using a
freedom of renormalization scheme (RS) choice in high-order PT. 
This provides a particular method of summation of asymptotic series 
related to
the IR renormalons in QCD.  For the method to work one should consider
higher order  PT corrections to the standard bubble-chain diagrams.

First we consider an example of how higher order PT contributions can
be used to give a definition of the IR renormalon
contributions.  Let the $\beta$-function of a theory in some
particular renormalization scheme  be summed into the expression \be
\beta(\alpha)=-{\alpha^2\over 1+k\alpha^2}, \quad k>0
\label{beta1}
\ee with the standard asymptotic $\beta$-function of the form $
\beta^{as}(\alpha)=-\alpha^2$. The renormalization group
equation for the running coupling constant $\alpha(z)$ \be
z{\partial\over\partial z}\alpha(z) =\beta(\alpha)
\label{rg2}
\ee has a solution \be \alpha(z)={-L_z+\sqrt{L_z^2+4k}\over 2k}, \quad
L_z=\ln{z\over \Lambda^2} .
\label{sol3}
\ee 
This solution has no singularities at positive  $z$.  However, the
asymptotic solution 
\be 
\alpha^{as}(z)=\frac{1}{\ln(z/\Lambda^2)} 
\ee
has a pole at $z=\Lambda^2$.  Thus, the particular
way of summing an infinite number of perturbative terms for the
$\beta$-function in eq.~(\ref{beta1}) has cured the Landau pole
problem in a sense that the coupling becomes smooth in the IR 
region and the integration  of the evolution trajectory can
be performed explicitly.  Let us stress that there are no
nonPT terms added but the freedom of choosing a
renormalization scheme for an infinite series defining the running
coupling was used instead.  This is, however, beyond the
formal framework of perturbation theory where only finite order
polynomials in the coupling constant are considered for
$\beta$-functions.  The problem of RS ambiguity  in
high orders of PT was discussed in refs.~\cite{stev,bro,dhar,pms} 
where several methods to reduce this
ambiguity have been proposed.  In contrast to previous approaches
where the main criterion  for a good scheme is a weak dependence on
the change of the scheme parameterization, in  the present paper we
use the RS freedom to find a scheme that  allows
one to perform the integration over the infrared region without
encountering a singularity.

Eq.~(\ref{beta1}) can be considered either as a pure PT
result in some particular renormalization scheme after an
infinite resummation or as an approximation of some exact
$\beta$-function  (as obtained from a numerical study on the lattice,
for instance) that might include nonperturbative terms as well.  The
essential point for us here is that the running coupling
obeying the renormalization group equation with such a
$\beta$-function can smoothly be extrapolated 
into the IR region.  The running coupling given in
eq.~(\ref{sol3})  has a correct asymptotic behavior at  $z\rightarrow
\infty$.  It is a regular expansion parameter for physical observables
that obey the dispersion relation because it has no Landau pole on the
physical cut.

The generalization of eq.~(\ref{beta1}) to QCD is now straightforward.
Keeping in mind that first two coefficients $\beta_0$ and $\beta_1$ of
the QCD $\beta$-function are renormalization scheme invariant one
could suggest a possible "improved" $\beta$-function as \be
\beta(\alpha)={-\beta_0\alpha^2-\beta_1\alpha^3\over
1+k(\beta_0\alpha^2+\beta_1\alpha^3)}, \quad k>0
\label{sol4}
\ee with the running coupling constant $\alpha(z)$ implicitly
determined by the equation \be \beta_0 \ln{z\over \Lambda^2}={1\over
\alpha(z)}- {\beta_1\over \beta_0}
\ln\left({\beta_0+\beta_1\alpha(z)\over\beta_1\alpha(z)}\right) -k
\beta_0\alpha(z)
\label{eq5}
\ee 
The $\beta$-function given in
eq.~(\ref{sol4})  is bounded at large $\alpha$ and eq.~(\ref{eq5}) has
a solution for $\alpha(z)$ that  is defined on the positive semi-axis
and has no singularities.

The solution given in eq.~(\ref{eq5}) can be rewritten
through an intermediate energy scale $\mu$ without using the
$\Lambda$ parameter (the $\Lambda$ parameter has no special meaning as
a position of the pole of the running coupling). 
For the sake of simplicity we do that for the
leading order expression given in eq.~(\ref{sol3})
and find 
\be 
\alpha(Q^2) ={2\alpha\over 1+\alpha L_Q-k\alpha^2
+\sqrt{(1+\alpha L_Q-k\alpha^2)^2+4k\alpha^2}}\, , 
\ee 
where
$L_Q=\ln(Q^2/\mu^2)$ and  $\alpha\equiv \alpha(\mu^2)$.  This form of
the evolution of the  coupling constant is a particular regularization
of the Landau singularity. This regularization introduces no imaginary
part  on the positive semi-axis (cf. ref.~\cite{Grunberg}).  Thus, in
the pure PT framework one can eliminate the Landau
pole of the coupling using the freedom of  the RS choice in high
orders. The smooth coupling requires no nonperturbative effects to
regularize the integrals in which it appears. The integral over the
small momenta determines a particular sum for a  series related to an
(extended) IR renormalon contribution.

The expression given in eq.~(\ref{sol4}) for the $\beta$-function is
not polynomial.  Still even for a two-loop $\beta$-function there can
exist  an IR fixed point that allows one to make an extrapolation to
small energies.  Indeed, for $n_f$ light flavors in QCD the first two
coefficients of the $\beta$-function read
\[
4\pi\beta_0=11-\frac{2}{3}n_f, \quad
(4\pi)^2\beta_1=102-\frac{38}{3}n_f
\]
and for
\[
\frac{153}{19}< n_f <\frac{33}{2}
\]
one has
\[
\beta_0>0, \quad \beta_1<0 \, .
\]
Therefore, for $n_f=9$, for instance,  there exists an infrared fixed
point and an extended  IR-renormalon asymptotic series  (generated
with a two-loop $\beta$-function) can be explicitly summed according
to our prescription, i.e.  in the second order approximation for the
evolution  of the coupling in QCD.

The coupling constant defined in a particular  renormalization scheme
is not an immediate physical quantity.  However, one can write a
formal definition  of the effective coupling through some physical
quantity. This definition can implicitly  include nonperturbative
terms. For instance, taking the expression for Adler function
\[
D_{e^+e^-}(Q^2)=Q^2\int{\frac{ R_{e^+e^{-}}(s)ds}{(s+Q^2)^2}}
\] 
one can define
\[
d^{as}(Q^2)=(D_{e^+e^-}(Q^2)-1)\pi\beta_0 =\left(\ln{Q^2\over
\Lambda^2}\right)^{-1}
\]
and \cite{analKr,pivtau}
\be
\label{reepi}
r_{e^+e^-}^{as}(s)={1\over \pi} \arctan\left({\pi\over
\ln{s\over\Lambda^2}}\right) \Theta(s)+\Theta(s+\Lambda^2)\Theta(-s).
\ee
The unphysical singularity  (a cut along a part of the negative
semi-axis) corresponds to the unphysical pole in $D(Q^2)$.
Subtracting this singularity in a minimal way we find the expression
\be 
d(Q^2) = {1\over \ln{(Q^2/ \Lambda^2)}}  - 
{ \Lambda^2\over Q^2-\Lambda^2}
\label{eq6}
\ee 
that is regular for positive $Q^2$
but has power corrections at large $Q^2$ and is not polynomial in
terms of the asymptotic charge
\[
d^{as}(Q^2) = \left(\ln{(Q^2/\Lambda^2)}\right)^{-1}\, .
\]
Indeed, for large $Q^2$ one finds \be d(Q^2)=d^{as}(Q^2)-{1\over
e^{1\over d^{as}(Q^2) }-1}= d^{as}(Q^2)- e^{-{1\over d^{as}(Q^2)}
}+\ldots
\label{eq7}
\ee 
The corresponding $\beta$-function written in terms of the
asymptotic charge $d^{as}\equiv \xi$ \be
\label{bfSh}
Q^2\frac{\partial}{\partial Q^2}d(Q^2) = -\xi^2+ e^{-\frac{1}{\xi}}(1-
e^{-\frac{1}{\xi}})^{-2}|_{\xi=d^{as}(Q^2)} 
\ee 
also has an explicit
nonperturbative term unlike  eq.~(\ref{sol4}).  It is unclear whether
the $\beta$-function in eq.~(\ref{bfSh}) can be rewritten in terms of the
full charge $d(Q^2)$ from eq.~(\ref{eq6}).

Thus, there are various possibilities to eliminate the leading order
pole and, thereby, to give an interpretation to the renormalon
series as an integral of the running coupling over the IR
region.  A pure perturbative possibility means that higher order 
PT corrections
make the evolution well-defined till the origin. This can be
achieved with both  finite-order and resummed expressions for the
$\beta$-function.  Other possibilities can be  nonperturbative ones when
the evolution is made smooth in the IR region 
by some nonperturbative procedure.

Within the IR-renormalon approach the power corrections emerge as   a
result of integration of some sub-blocks of a whole diagram over the
region of small momenta in the phase space of the diagram.  The main
generating sub-block  in the IR-renormalon approximation is taken to be
the running coupling.  However, for defining the IR
contributions there is no need to know  the point-by-point
evolution of the coupling in the IR domain.  One has to define some
integrals only.
Finding only integrals is less demanding than 
determining the point-by-point evolution
of the coupling in the IR domain.  For phenomenological applications
the information about a particular way of extrapolation of the
coupling constant evolution into the IR domain can be encoded through
just a nonPT parameter describing the leading order nonPT
correction.  In other words any particular prescription of summing the
asymptotic series generated by the (extended) IR-renormalon chains
corresponds to a numerical value of this parameter.  For instance, the
parameter $k$ in eq.~(\ref{beta1}) describes a way of summing the
IR-renormalon asymptotic series by using 
higher order corrections and changing
the renormalization scheme in which the evolution is described.  For
phenomenological applications, however, it is convenient  to retain a
contact with the leading order asymptotic charge.  By choosing the
principal value (PV) prescription  for the regularization of integrals of
the asymptotic charge over the IR region one obtains a particular way
of resumming the IR-renormalon contributions.  This way is not
unique. The residual freedom is parameterized  by a nonPT quantity
which is IR sensitive only \cite{pen}.  The parameterization is
written in the form 
\be 
\beta_0\alpha^{eff}(s)
=\beta_0\alpha_s(s)|_{\rm PV} +A\Lambda^2\delta(s)
+B\Lambda^4\delta'(s)+\ldots
\label{eq8}
\ee 
The normalization is chosen such that  for large $s$ 
\be
\alpha_s(s)=\frac{1}{\beta_0 \ln (s/\Lambda^2)}\, .
\label{norm}
\ee 
Eq.~(\ref{eq8}) is understood in a sense of distributions  and
$\alpha_s(s)|_{\rm PV}$ is the asymptotic charge containing the pole
regularized by taking the principal value.  Such a distribution is a
natural generalization of the classical solution to the
renormalization group equation with the asymptotic $\beta$-function
$\beta^{as}(x)=-\beta_0 x^2$.  The generalized solution
$\alpha_s(s)|_{\rm PV}$ is not unique and allows some arbitrariness at
the point $s=\Lambda^2$ that should be a series of $\delta$-function
and its derivatives.  Quantities $A,B,...$ parameterize this
arbitrariness.  We have chosen  the support of "nonperturbative terms"
at $s=0$ instead of $s=\Lambda^2$ which suffices for the 
discussion in the present paper. However, if eq.~(\ref{eq8})
is integrated with a weight function that is singular in the origin 
one has to retain the exact location of the support 
$s=\Lambda^2$ of the "nonperturbative terms".
It is necessary, for instance, in case of finding renormalon
corrections to the pole mass of a heavy quark.
Note that multiplication is not defined for
distributions  therefore the non-leading PT corrections are supposed to be
included in the leading order term in eq.~(\ref{eq8})
that results in the substitution 
$\Lambda_{\MSsch}\to \Lambda_{\rm eff}$ (e.g. \cite{gr,kras,gg}).  
The integration of expression in
eq.~(\ref{eq8}) over the IR region with some numerical values for the
parameters $A$ and $B$ should give the same result as an integration
of  the solution to eq.~(\ref{sol4}) with some numerical value for the
parameter $k$. For a while we retain two parameters in eq.~(\ref{eq8}).

Now we consider several models for $\alpha^{eff}(s)$ and  compute
numerical values for parameters $A$ and $B$. In the spirit of the
operator product  expansion $A=0$ (there are no gauge invariant
operators of dimensionality two) and $B\sim \langle G^2\rangle$, where
$\langle G^2\rangle$ is the gluon condensate.  With the constraint
$A=0$ the parameters $B$ is related  to the parameter $k$ from
eq.~(\ref{sol4}).  Within such a model one expects that  the pattern
of extrapolation of the running coupling to the infrared region can be
found by comparing the numerical value for the  parameter $B$ with 
that of the gluon condensate.  
The gluon condensate is connected with  the violation of
dilatation symmetry which is closely related to dimensional
transmutation and appearance of the scale $\Lambda$.
Note that quark condensates are related to chiral
symmetry breaking and we do not consider them within our approach.  

Assuming a universality of extrapolation
of the running coupling into the IR region we fix the parameter $B$ from
comparison with OPE.  Taking the process of $e^+e^-$ annihilation
where the operator product expansion is applicable we find the
relation between the $B$ parameter and the gluon condensate.  From the
expression for Adler's function at large Euclidean $Q^2$ 
\be
D_{e^+e^-}(Q^2)=1+\frac{\al_s^{e^+e^-}(Q)}{\pi} +\frac{\langle g_s^2
G^2\rangle}{6 Q^4}
\label{adlfun}
\ee 
one obtains the (naive) asymptotic spectral density in the leading order 
\be
\label{spectfun}
R_{e^+e^-}(s)=1+\frac{\al_s^{e^+e^-}(s)}{\pi} +\frac{\langle g_s^2
G^2\rangle}{12}\delta'(s) 
\ee 
with  
\be
\label{norm2}
\alpha_s^{e^+e^-}(s)=\frac{1}{\beta_0 \ln (s/\Lambda_{e^+e^-}^2)}\, .
\ee 
Note that $\pi^2$-corrections are ignored
(cf eq.~(\ref{reepi})).
Assuming the interpretation of the asymptotic charge
$\alpha_s^{e^+e^-}(s)$ in the above
equations as a distribution on the semi-axis $0<s<\infty$ with PV
regularization we have  
\be 
\al_s^{eff}(s)= \al_s^{e^+e^-}(s)_{\rm PV}
+\frac{\pi}{12}\langle g_s^2 G^2\rangle\delta^{\prime}(s)\, .
\label{spectfun1}
\ee 
This equation fixes a particular extrapolation of the coupling 
evolution into the IR domain.
The chosen extrapolation is represented through 
the PV regularized asymptotic charge and the gluon condensate
contribution which has support only in the IR domain. 
This extrapolation is coordinated  with OPE for the two-point
correlator in $e^+e^-$ annihilation. 
Comparing eq.~(\ref{spectfun1}) with eq.~(\ref{eq8}) one finds
for three light quarks
\[
16 B\Lambda_{e^+e^-}^4 =3\langle g_s^2 G^2\rangle
\]
that translates into the following expression for the scale parameter
$\Lambda_{e^+e^-}$ through the gluon condensate
\[
\Lambda_{e^+e^-} =\frac{1}{2}\left(\frac{3\langle g_s^2
G^2\rangle}{B}\right)^{\frac{1}{4}} .
\]
For the standard numerical value 
$\langle g_s^2 G^2\rangle = 0.474~{\rm GeV}^4$ 
we find
\be
\label{fff}
\Lambda_{e^+e^-}=\frac{546}{B^{\frac{1}{4}}}~{\rm MeV} \, .
\ee
In $e^+e^-$ annihilation the parameter $\Lambda_{e^+e^-}$ is
an effective scale of the process \cite{kras,gr}.
The use of the effective scale
allows one to absorb the next-to-leading order PT correction
into $\alpha_s^{e^+e^-}(s)$.
The  numerical value of the effective scale is
rather sensitive to the process and can characterize the scale 
where PT is violated and nonPT term are important numerically
\cite{gg,spinn}.
To find the relation between $\Lambda_{e^+e^-}$ and the standard 
parameter $\Lambda_{\overline{\rm MS}}$
we consider the first perturbative correction 
to the coupling constant $\alpha_s^{e^+e^-}(s)$ in eq.~(\ref{adlfun})
\be
\frac{\alpha_s^{e^+e^-}(\mu)}{\pi}
=\frac{\alpha_s^{\overline{\rm MS}}(\mu)}{\pi}
+k_1\left(\frac{\alpha_s^{\overline{\rm MS}}(\mu)}{\pi}\right)^2
+O(\alpha_s^3)
\ee
where 
\[
k_1=\frac{365}{24}-11\zeta(3)
+n_f\left(-\frac{11}{12}+\frac{2}{3} \zeta(3)\right).
\] 
Here $\zeta(z)$ is Riemann $\zeta$-function.
For the standard QCD with $n_f=3$ we find numerically
\be
\frac{\alpha_s^{e^+e^-}(\mu)}{\pi}
=\frac{\alpha_s^{\overline{\rm MS}}(\mu)}{\pi}
+1.64\left(\frac{\alpha_s^{\overline{\rm MS}}(\mu)}{\pi}\right)^2
+O(\alpha_s^3)
\ee
that leads to 
\be
\label{link}
\Lambda_{e^+e^-}=1.44~\Lambda_{\overline{\rm MS}}\, . 
\ee
Using eqs.~(\ref{fff},\ref{link}) one obtains a numerical prediction
\be
\Lambda_{\overline{\rm MS}}
=380~B^{-\frac{1}{4}}~{\rm MeV} \, .
\label{pred}
\ee
For the present experimental value of 
$\Lambda_{\overline{\rm MS}}^{\rm exp}$
extracted from the data on the $\tau$-lepton decay \cite{pivKraj}
\be
\Lambda_{\overline{\rm MS}}^{\rm exp}=349\pm 61~{\rm MeV}
\label{expla}
\ee
one finds
\be
B=\left(\frac{380}{349\pm 61}\right)^4=0.74\div 3.03\, .
\label{Bpar}
\ee
Below we consider several explicit 
models of extrapolation of the coupling
evolution into the IR region
that allow one to determine a numerical value for the parameter $B$.
We find that the 
numerical value for the parameter $B$ obtained from these models
agrees with the result given in eq.~(\ref{Bpar}).
For the sake of technical simplicity 
we do not use sophisticated expressions for running coupling 
in the IR region stemming from RS redefinition.
We limit ourselves to models with rather a similar 
pattern of the IR behavior but 
expressible in simple analytical terms
that makes the numerical analysis 
more transparent.

{\bf Model 1.}
 
\noindent Switching the interaction off completely at small momenta:
\[
\alpha^{eff}(z)=\alpha_s(z)\Theta(z-a\Lambda^2).
\]
The system of equations for determining $A$ and $B$ is
\cite{FESR}
\[
li(a)+A=0, \quad li(a^2)-B=0 ,
\]
where $li(a)$ is a special function
\[
li(a)=\int^a_0{dt\over \ln (t)}
\]
with the PV (principal value)
prescription for the integration across the pole at $t=1$ 
for real positive $a>1$ \cite{GrR}.  
The numerical solution to the system of equations 
with an additional constraint $A=0$ reads $a=1.45$,
$B=li(2.1)=1.19$. We retain only three significant figures.$\bullet$

{\bf Model 2.}

\noindent Freezing the running coupling at small
momenta (e.g. \cite{kraspiv})
\[
\alpha^{eff}(s)=\alpha_s(a)\Theta(a\Lambda^2-s)
+\alpha_s(s)\Theta(s-a\Lambda^2) \, .
\]
The system of equations for determining $A$ and $B$ is
\[
{a\over \ln (a)}=li(a)+A,
\quad {a^2\over 2\ln (a)}=li(a^2)-B.
\]
The  numerical solution with the constraint $A=0$ reads
$a=3.85$, $B=2.6$.$\bullet$

{\bf Model 3.}

\noindent Regularization of the low energy behavior 
by the minimal subtraction of the pole singularity 
\[
\beta_0\alpha^{eff}(s)={1\over \ln(s/\Lambda^2)}
-{\Lambda^2\over s-\Lambda^2}.
\]
The system of equations to determine the parameters $A$ and $B$ 
has the form
\[
li(a)-\ln(a-1)=li(a)+A, \quad li(a^2)-a-\ln(a-1)=li(a^2)-B.
\]
The solution with $A=0$ is $a=2$, $B=2$.
Note that this model has recently got some special attention \cite{shir}.
$\bullet$

{\bf Model 4.}

\noindent Renormalization scheme modification of the
$\beta$-function and smooth extrapolation of the evolution  to small
momenta. The model is given by the pattern of running  presented in
eq.~(\ref{sol3}). The solution depends on the numerical value for the
parameter $k$ entering the expression for the $\beta$-function in
eq.~(\ref{beta1}).   For $k=2$ the numerical solution is $a=3.0$,
$B=3.0$ which  is close to that of Models 2,3. For larger $k$ the
solution  is rather stable. For smaller values of $k$ there may be no
solution at all ($k$=1).  This shows that the anzats chosen in
eqs.~(\ref{eq8},\ref{spectfun1}) restricts the possible extrapolations
of the effective coupling into the IR region in Minkowskian domain and
can be used to fix the numerical value for the $B$ parameter through
the gluon condensate.$\bullet$

Thus, the models of reasonably soft extrapolation  of the running to
small momenta give the value for the parameter $B$ that agrees
with the estimate given in eq.~(\ref{Bpar}).   
Assuming that the numerical value for $B$ has been found independently
we can find a  numerical value for 
the QCD scale parameter from eq.~(\ref{pred}).
Using the results of Models 1-3
$B=1.2-2.6$ or $B^{\frac{1}{4}}=1.05-1.27$ one
obtains   $\Lambda_{\overline{\rm MS}}=300\div 362~{\rm MeV}$, or
$\Lambda_{\overline{\rm MS}}=331\pm 31~{\rm MeV}$  that is in a
reasonable agreement with the present data.  In general, our result
means that the numerical values of the gluon condensate and
$\Lambda_{\overline{\rm MS}}$ are compatible with each other
for the smooth continuation of the running coupling into  the IR
region.  The gluon condensate is interpreted as a quantity that
corrects the PV-regularized  leading-order running coupling at small
momenta  (in a sense of distributions) to provide a smooth
continuation of the evolution into the IR domain.

To conclude, we have exploited the RS freedom  in high orders of PT to
show that contributions of  some infinite subsets of diagrams that are
represented  as integrals of the running coupling  over the IR
region can explicitly be summed in a generalized way through the
proper definition of the integral.  The lack of any parameter or
sensible strict criterion for choosing a particular set of diagrams
that dominate high-order behavior beyond the leading order expression
for the running coupling allows one to use the RS freedom  for
interpretation of the IR-renormalon singularities.  If in a particular
renormalization scheme there are no  singularities  of the running
coupling at small momenta,  this scheme provides a particular recipe
for the resummation of the asymptotic series related to  the IR
renormalons.  As for phenomenological applications, the models with
some extrapolation of the running coupling into the infrared region
are used for practical calculation.  Our results show that with an
extrapolation chosen one cannot freely add the standard contribution
of  the gluon condensate to take into account nonperturbative effects
and "improve" the computation -- this contribution  must be
coordinated with the extrapolation of the coupling.


\begin{thebibliography}{99}
\bibitem{lip}L.N. Lipatov, Zh. Eksp. Teor. Fiz. 72, 411 (1977).
\bibitem{lip1}A.P. Bukhvostov and L.N. Lipatov,  
Zh. Eksp. Teor. Fiz. 73, 1658 (1977).
\bibitem{JZ}J. Zinn-Justin, Quantum field theory and critical
phenomena, Third edition, Oxford, 1996.
\bibitem{thooft}
G.'t Hooft, The Whys of Subnuclear Physics, Proceedings of the
15th Inter.School on Subnuclear Physics, Erice, Sicily, 1977, edited
by A.Zichichi (Plenum Press, New York, 1979), p.943.
\bibitem{lautrup}
B. Lautrup, Phys. Lett. B69, 109 (1977). 
\bibitem{hardy}G.H. Hardy, Divergent series, Oxford 1973.
\bibitem{Parisi}
G. Parisi, Phys. Lett. 76B, 65 (1978);
Nucl. Phys. B150, 163 (1979).
\bibitem{Mueller}
A.H. Mueller, Nucl. Phys. B250, 327 (1985).
\bibitem{bal}
I.I. Balitsky, Phys. Lett. B273, 282 (1991).
\bibitem{zakh}
V.I. Zakharov, Nucl. Phys. B385, 452 (1992).
\bibitem{pivKraj0}
J.G. Korner, F. Krajewski and A.A. Pivovarov, 
Eur. Phys. J.  C12, 461 (2000); \\
{\em ibid} C14, 123 (2000).
\bibitem{politzer}H.D. Politzer, Nucl. Phys. B117, 397 (1976).
\bibitem{SVZcharm} 
A. Vainshtein, V. Zakharov, M. Shifman,
JETP Lett. 27, 59 (1978).
\bibitem{Wilson}
K.G. Wilson, Phys. Rev. 179, 1399 (1969). 
\bibitem{david}
F. David, Nucl. Phys. B209, 433 (1982).
\bibitem{tmf}A.A. Pivovarov, A.N. Tavkhelidze and V.F. Tokarev, \\
Theor. Math. Phys. 60, 765 (1985).
\bibitem{SVZ} 
M.A. Shifman, A.I. Vainshtein, V.I. Zakharov, 
Nucl. Phys. B147, 385 (1979).
\bibitem{Coquer}
R. Coquereaux, Phys.~Rev. D23, 2276 (1981).
\bibitem{be0}
D.J. Broadhurst and A.G. Grozin,  
Phys. Rev. D52, 4082 (1995);\\
M. Beneke, V.Braun,  Phys. Lett. B348, 513 (1995);\\
C.N. Lovett-Turner, C.J. Maxwell, Nucl. Phys.
B452, 188 (1995).
\bibitem{kor}
G.P. Korchemsky and  G. Sterman,  
Nucl. Phys. B437, 415 (1995).
\bibitem{khal}L.D.~Landau, A.A.~Abrikosov, I.M.~Khalatnikov,  \\
Doklad. Akad. Nauk SSSR 95, 773 (1954). 
\bibitem{RGeqGM}
M. Gell-Mann, F.E. Low, Phys.Rev. 95, 1300 (1954).
\bibitem{RGeqBS}
N.N. Bogoliubov, D.V. Shirkov, Doklad. Akad. Nauk SSSR 103,  391 (1955)\\
Nuovo Cim. 3, 845 (1956).
\bibitem{aavlad}A.A. Vladimirov, Teor. Mat. Fiz. 43, 210 (1980).
\bibitem{krpiv}N.V. Krasnikov and A.A. Pivovarov,
INR-903-95, Oct 1995. 9pp., to be published
in Yad. Fiz. [hep-ph/9512213]; \\
INR-925-96, Jul 1996. 8pp. 
Talk given at 10th International Conference on Problems of Quantum 
Field Theory, Alushta, Ukraine, 13-17 May 1996. 
Proceedings of 10th International Conference on Problems of Quantum Field
Theory, Eds. D.V.Shirkov, D.I.Kazakov, A.A.Vladimirov, 
Dubna 1996, p.178 - 183. [hep-ph/9607247];\\
Mod. Phys. Lett. A11, 835 (1996).
\bibitem{kraspiv}
N.V. Krasnikov and A.A. Pivovarov,
Sov. Phys. J. 25, 55 (1982).
\bibitem{beneke}M. Beneke, Phys. Rept. 317, 1 (1999).
\bibitem{HQm}
I.I. Bigi, M.A. Shifman, N.G. Uraltsev, A.I. Vainshtein,
Phys. Rev. D50, 2234 (1994).
\bibitem{silv}S.V. Faleev, P.G. Silvestrov,
Nucl. Phys. B507, 379 (1997).
\bibitem{suslov}I.M. Suslov,   
Zh. Eksp. Teor. Fiz. 116, 369 (1999). 
\bibitem{hoang}A.~Hoang {\it et al.},
Eur. Phys. J. direct C3, 1 (2000).
\bibitem{pp}A.A. Penin and A.A. Pivovarov, 
Phys. Lett.  B435, 413 (1998);\\
Nucl. Phys.  B549, 217 (1999);
MZ-TH-98-61, Dec 1998. 41pp., to be published in Yad. Fiz.
[hep-ph/9904278].
\bibitem{stev} P.M. Stevenson,  
Phys. Rev. D23, 2916 (1981).
\bibitem{bro}J. Brodsky, G.P. Lepage, P.B. Mackenzie,
Phys. Rev. D28, 228 (1983).
\bibitem{dhar}
A. Dhar and V.Gupta, Phys. Rev. D29, 2822 (1984).
\bibitem{pms}
A.C. Mattingly and  P.M. Stevenson,  Phys. Rev. D49, 437 (1994).
\bibitem{Grunberg}
G. Grunberg, Phys. Lett. B349, 469 (1995).
\bibitem{analKr}N.V. Krasnikov, A.A. Pivovarov, 
Phys. Lett. 116B, 168 (1982).
\bibitem{pivtau}A.A. Pivovarov, Sov. J. Nucl. Phys. 54, 676 (1991);
Z.Phys. C53, 461 (1992); \\
Nuovo Cim. 105A, 813 (1992).
\bibitem{pen}A.A. Penin and A.A. Pivovarov,
Phys. Lett. B357, 427 (1995); {\em ibid} B367, 342 (1996);
{\em ibid} B401, 294 (1997); Phys. Atom. Nucl. 60, 2056 (1997).
\bibitem{gr}G. Grunberg,  Phys. Lett. B90, 70 (1980).
\bibitem{kras}N.V. Krasnikov, Nucl. Phys. B192, 497 (1981).
\bibitem{gg}A.L. Kataev,  N.V. Krasnikov and A.A. Pivovarov, 
Phys. Lett. B107, 115 (1981);
Nucl. Phys. B198, 508 (1982); {\em ibid} B490, 505 (1997).
\bibitem{spinn}A.A. Pivovarov and E.N. Popov,
Phys. Lett. B205, 79 (1988);\\
Sov. J. Nucl. Phys. 49, 693 (1989).
\bibitem{pivKraj}J.G. K\"orner, F. Krajewski and A.A. Pivovarov, 
MZ-TH-00-03, Feb 2000. 11pp., \\
Phys. Rev. D, to be published 
[hep-ph/0002166]. 
\bibitem{FESR}N.V. Krasnikov, A.A. Pivovarov, N.N. Tavkhelidze,\\
JETP Lett. 36, 333 (1982); Z.Phys. C19, 301 (1983).
\bibitem{GrR}I.S. Gradshtein, I.M. Ryzhik,
Tables of integrals, sums and series.
Moscow, 1951.
\bibitem{shir}D.V. Shirkov and I.L. Solovtsov,  
Phys. Rev. Lett. 79, 1209 (1997).












\end{thebibliography}
\end{document}